# The 1.5 Ms Observing Campaign on IRAS 13224−3809: X-ray Spectral Analysis I.

J. Jiang (姜嘉陈) [1][*], M. L. Parker[13,1], A. C. Fabian[1], W. N. Alston[1], D. J. K. Buisson[1], E. M. Cackett[5], C.-Y. Chiang[5], T. Dauser[6], L. C. Gallo[3], J. A. García[9,6], F. A. Harrison[9], A. M. Lohfink[8], B. De Marco[2], E. Kara[11], J. M. Miller[12], G. Miniutti[4], C. Pinto[1], D. J. Walton[1], and D. R. Wilkins[7]

[1] *Institute of Astronomy, Univeristy of Cambridge, Madingley Road, CB3 0HA Cambridge, UK*
[2] *Nicolaus Copernicus Astronomical Center, Polish Academy of Sciences, Bartycka 18, PL-00-716 Warsaw, Poland*
[3] *Department of Astronomy and Physics, Saint Mary's University, 923 Robie Street, Halifax, NS, B3H 3C3, Canada*
[4] *Centro de Astrobiología (INTA–CSIC), Dep. de Astrofísica, ESAC campus, 28692 Villanueva de la Cañada, Spain*
[5] *Department of Physics and Astronomy, Wayne State University, 666 W, Hancock Street, Detroit, MI 48201, USA*
[6] *Dr Karl Remeis-Observatory and Erlangen Centre for Astroparticle Physics, Sternwartstr. 7, D-96049 Bamberg, Germany*
[7] *Kavli Institute of Particle Astrophysics and Cosmology, Standford University, 452 Lomita Mall, Standford, CA 94305, USA*
[8] *Department of Physics, Montana State University, Bozeman, MT 59717-3840, USA*
[9] *Cahill Center for Astronomy and Astrophysics, California Institute of Technology, Pasadena, CA 91125, USA*
[10] *Harvard-Smithsonian Center for Astrophysics, Cambridge, MA 02138, USA*
[11] *Department of Astronomy, University of Maryland, College Park, MD 20742-2421, USA*
[12] *University of Michigan, Department of Astronomy, 1085 S. University, Ann Arbor, MI 48109, USA*
[13] *European Space Agency (ESA), European Space Astronomy Centre (ESAC), E-28691 Villanueva de la Cañada, Madrid, Spain*

**ABSTRACT**

We present a detailed spectral analysis of the recent 1.5 Ms *XMM-Newton* observing campaign on the narrow line Seyfert 1 galaxy IRAS 13224−3809, taken simultaneously with 500 ks of *NuSTAR* data. The X-ray lightcurve shows three flux peaks, registering at about 100 times the minimum flux seen during the campaign, and rapid variability with a time scale of kiloseconds. The spectra are well fit with a primary powerlaw continuum, two relativistic-blurred reflection components from the inner accretion disk with very high iron abundance, and a simple blackbody-shaped model for the remaining soft excess. The spectral variability is dominated by the power law continuum from a corona region within a few gravitational radii from the black hole. Additionally, blueshifted Ne X, Mg XII, Si XIV and S XVI absorption lines are identified in the stacked low-flux spectrum, confirming the presence of a highly ionized outflow with velocity up to $v = 0.267$ and $0.225$ c. We fit the absorption features with `xstar` models and find a relatively constant velocity outflow through the whole observation. Finally, we replace the `bbody` and supersolar abundance reflection models by fitting the soft excess successfully with the extended reflection model `relxillD`, which allows for higher densities than the standard `relxill` model. This returns a disk electron density $n_e > 10^{18.7}$ cm$^{-3}$ and lowers the iron abundance from $Z_{Fe} = 24^{+3}_{-4} Z_\odot$ with $n_e \equiv 10^{15}$ cm$^{-3}$ to $Z_{Fe} = 6.6^{+0.8}_{-2.1} Z_\odot$.

# 1 INTRODUCTION

## 1.1 Accretion Disk-Corona for AGN

The primary X-ray emission from black holes can be described by a powerlaw continuum with a high energy cutoff,

---
[*] E-mail: jj447@cam.ac.uk

© 2018 The Authors

often explained by inverse Comptonization of the thermal disk photons in a coronal region (e.g. Haardt & Maraschi 1993). The corona is known to be compact, though its exact nature still remains unknown. Advanced imaging and timing analyses show the X-ray emitting regions are highly compact and only a few gravitational radii from the black hole (Reis & Miller 2013, for review). For example, the discovery of a soft X-ray reverberation lag of 30 s in 1H0707-495, a narrow line Seyfert 1 galaxy (NLS1), has indicated the X-ray emitting region is very compact (Fabian et al. 2009), which will be discussed more later. Moreover, in the microlensed galaxy the microlensing duration in the X-ray band is much shorter than the UV/optical band, which indicates a much smaller X-ray emitting region than the optical emitting region (e.g. Morgan et al. 2008; Chartas et al. 2017). Some progress on modeling the disk emissivity profile with different compact corona geometries has also been made (e.g. Wilkins et al. 2015; Wilkins & Gallo 2015), and is consistent with a compact coronal region, close to the event horizon. The compact corona also agrees with the predictions of the Comptonization model with magnetic reconnection (e.g. Merloni & Fabian 2001), where the corona is described as a region of smooth magnetic field with increasing strengths towards small radii. A coronal geometry in which the continuum originates from a small region on the spin axis at height $h$ above the central black hole can describe X-ray data from many luminous accreting black holes well (e.g. MCG-6-30-15, Fabian & Vaughan 2003; Miniutti et al. 2003; Vaughan & Fabian 2004). Such a geometry is usually called the lamp-post geometry.

The reprocessing of the coronal radiation by the colder material in the disk produces a hump above 20 keV and a series of atomic lines, most notably the strong Fe K$\alpha$ emission line at 6.4 keV. These features are referred to as the disk reflection component. The interaction between the primary powerlaw photons and the disk material can produce both emission, including fluorescence lines and recombination continuum, and absorption edges (George & Fabian 1991; Ross & Fabian 2005; García & Kallman 2010). X-ray reflection off the inner part of the accretion disk is highly affected by strong gravitational effects including gravitational redshift and Doppler effects (e.g. Fabian et al. 1989; Reynolds & Nowak 2003), which can offer information on the geometry of the corona and the spin of the central black hole. For example, relativistic broad Fe line features have been detected in the reflection spectra of many AGN sources, such as MCG-6-30-15 (e.g. Tanaka et al. 1995; Wilms et al. 2001; Fabian & Vaughan 2003; Marinucci et al. 2014b), 1H0707-495 (e.g. Fabian et al. 2004, 2009), NGC 1365 (e.g. Risaliti et al. 2013; Walton et al. 2014), Mrk 335 (e.g. Larsson et al. 2008; Gallo et al. 2013; Walton et al. 2013; Parker et al. 2014), IRAS 00521-7054 (e.g. Tan et al. 2012), NGC 3783 (e.g. Brenneman et al. 2011), Swift J2127.4+5654 (e.g. Miniutti et al. 2009; Marinucci et al. 2014a). Recent work on AGN X-ray variability has shown time lags where the soft excess and Fe K$\alpha$ line lag behind the power law continuum. Evidence of lags of < 100 s, referred to as reverberation lags, has been detected in AGNs (e.g. Fabian et al. 2009; Zoghbi et al. 2010, 2011, 2012; De Marco et al. 2013; Kara et al. 2013a, 2016). Lags of a few milliseconds have also been detected in X-ray binaries (e.g. De Marco et al. 2015; De Marco & Ponti 2016; De Marco et al. 2017), indicating similar processes are driving the spectral variability in both classes of black hole.

### 1.2 IRAS 13224−3809

IRAS 13224−3809 ($z = 0.066$, Allen et al. 1991) is classified as a narrow line Seyfert-1 galaxy hosting a supermassive black hole ($M = 10^6 - 10^7 \, M_\odot$, Zhou & Wang 2005). It has been studied in multiple bands previously. It was identified by Boller et al. (1997) as a radio quiet source, with no clear jet emission yet observed (1.4GHz flux of 5.4 mJy, Feain et al. 2009). UV continuum variability of 24 per cent in three years has been observed (Rodriguez-Pascual et al. 1997). Leighly (2001) found asymmetric high ionization emission lines in the HST UV spectrum, indicating evidence of an outflow. $Ly_\alpha$ line variability has been detected with variation in the line profile and flux level (Mas-Hesse et al. 1994). Though there is no significant rapid optical variability detected (Young et al. 1999), it exhibits extreme and rapid variability in the X-ray band on very short time scales of hundreds of seconds (Boller et al. 1997; Gallo et al. 2004; Fabian et al. 2013)

IRAS 13224−3809 was observed by the *XMM-Newton* satellite (Jansen et al. 2001) in 2002 and 2011, showing a steep spectrum (e.g. $\Gamma = 2.5 - 2.7$, Boller et al. 2003; Fabian et al. 2013) with an obvious soft excess below 1.5 keV and a sharp spectral drop around 8 keV (Boller et al. 2003; Ponti et al. 2010; Fabian et al. 2013; Chiang et al. 2015). The spectrum shows very strong Fe K and Fe L emission lines (Ponti et al. 2010; Fabian et al. 2013) which are explained as the indication of reflection off the inner accretion disk around a rapidly rotating black hole (e.g. $a_* = 0.988 \pm 0.001$, Fabian et al. 2004, 2013). By analyzing the disk emissivity profile in the 500 ks long *XMM-Newton* observation in 2011, Fabian et al. (2013) found that the corona is located within a few gravitational radii. Chiang et al. (2015) fitted the soft excess in the EPIC spectra successfully with the combination of the reflection model `reflionx` (Ross & Fabian 2005) and a black body. The RGS spectra do not exhibit any evidence for absorption features produced by a warm absorber (Pinto et al. 2017b) or partial covering clouds (Chiang et al. 2015), and there is no evidence for absorption in the UV spectra (Leighly & Moore 2004).

A soft (0.3–1 keV) lag behind the 1–4 keV band was found by Ponti et al. (2010), similar to the lag detected in 1H0707−495 (Fabian et al. 2009; Zoghbi et al. 2010), though the significance of the lags is limited due to the short duration of the *XMM-Newton* observation in 2002. A more detailed timing study of the longer observation in 2011 found a strong Fe K feature in the lag-energy spectrum and suggested that the frequency and the amplitude of the lag varies in high flux intervals and quiescent periods (Kara et al. 2013b). These changes in the lag properties with flux are thought to be due to a more compact disk/corona system. Chainakun et al. (2016) simulated the time-averaged model and lag-energy model in the lamp-post scenario and the model also supports a reflection origin for the soft excess of IRAS 13224−3809 and an X-ray source very close to the central black hole ($h \approx 2 \, R_g$).

IRAS 13224−3809 was recently observed with a *XMM-Newton* quasi-simultaneous 1.5 Ms and *NuSTAR* (Harrison et al. 2013) observing campaign in 2016 (P.I. A. C. Fabian).

In Parker et al. (2017b), we reported the discovery of a series of strongly blueshifted ($v = 0.236 \pm 0.006\,c$) absorption lines from O, Ne, and Fe in the *XMM-Newton* EPIC-pn and RGS spectra, indicating the presence of an ultra-fast outflow (UFO). These features in IRAS 13224−3809 were found to be strongly dependent on the source flux, varying on timescales as short as a few kiloseconds. Similar outflows have been detected in many sources by studying the blue-shifted Fe K-shell absorption lines in the 7–10 keV energy band (e.g. Tombesi et al. 2010), but none has ever been seen to vary as rapidly, or to have strong correlation with flux (Parker et al. 2017a; Pinto et al. 2017b). The flux dependence can be explained as the increased source flux ionizing the outflowing gas, saturating the absorption lines. Alternatively, the absorption features could be produced in a layer on top of the accretion disk, where these relativistic velocities occur naturally. In this case, the change in strength of the absorption is due to the change in reflection fraction with flux, as only the reflected emission is absorbed (Gallo & Fabian 2011). This model has been successfully applied to PG 1211+143 (Gallo & Fabian 2013). Following on from the detection of the UFO, in Parker et al. (2017a) we showed that the Fe XXV/XXVI absorption features are strongly present in simple variability spectra, along with the Mg XII, Si XIV, S XVI, Ar XVIII and Ca XX Ly $\alpha$ lines. Again, all of these features are strongly flux dependent and extremely rapidly variable.

IRAS 13224−3809 is well known as a twin to another bright, rapidly variable NLS1: 1H 0707-495, which also shows very strong Fe L and Fe K emission in the *XMM-Newton* spectrum (Fabian et al. 2009). A reverberation lag of 30 s from the bright Fe L emission line was found by Fabian et al. (2009); Zoghbi et al. (2010). Kara et al. (2013a) found a low-frequency hard lag corresponding to the disk fluctuation propagation, and a high-frequency soft lag corresponding to the time delay between the coronal emission and the disk reflection in a 1.3 Ms observing campaign. In addition, Dauser et al. (2012) first found a highly ionized outflowing wind with changing velocity from 0.11 c to 0.18 c by studying the 500 ks long *XMM-Newton* and 120 ks *Chandra* quasi-simultaneous observations. Boller et al. (2002) found a sharp spectral drop around 7 keV in the *XMM-Newton* spectra, suggesting a neutral Fe K edge. Hagino et al. (2016) interpreted the absorption feature above 7 keV with a disk wind model. More data analysis on the ionized outflow in this source will be published in near future.

In this work, we analyze the 1.5 Ms *XMM-Newton* observing campaign of IRAS 13224−3809 using broad band spectroscopy. We focus on the spectral characteristics of the AGN, and compare our results with previous studies. In Section 2, we briefly describe the data reduction and briefly introduce the X-ray variability shown in the light curve; in Section 3, we focus on the stacked spectral analysis; in Section 4, we study the spectral differences between different flux levels; in Section 5, we divide the whole exposure into 12 slices and study the spectral variability. A detailed timing analysis, including further reverberation lag studies will be presented in future work (Alston et al. submitted) and the RGS spectral analysis is in an independent paper (Pinto et al. 2017a).

## 2 DATA REDUCTION AND LIGHT CURVES

Our data analysis is conducted using all the data from the 2016 *XMM-Newton* observing campaign. A list of the observations used in this paper for spectral analysis, with a total exposure of $\approx 1.5\,\mathrm{Ms}$, is given in Table 1. A list of quasi-simultaneous *NuSTAR* observations used in this work is shown in Table 2.

### 2.1 *XMM-Newton* Data Reduction

We reduce the *XMM-Newton* data using V15.0.0 of the *XMM-Newton* Science Analysis System (SAS) software package and calibration files (ccf) v.20160201. First, the clean calibrated event lists are created by running EMPROC (for EPIC-MOS data) and EPPROC (for EPIC-pn data). Good time intervals are selected by removing intervals dominated by flaring particle background, defined as intervals where the single event (PATTERN=0) count rate in the >10 keV band is larger than 0.35 counts s$^{-1}$ for EPIC-MOS data and that in the 10–12 keV band larger than 0.4 counts s$^{-1}$ for EPIC-pn data. By running the EVSELECT task, we select single and double events for EPIC-MOS (PATTERN<=12) and EPIC-pn (PATTERN<=4, FLAG==0) source event lists from a circular source region with radius of 35 arcsec. An annulus shaped source region with an inner radius of 5 arcsec is chosen to reduce pile-up if necessary for the observations with a very high count rate (see the last column of Table 1). The inner radius is chosen by running `epatplot`, with which the observed/model singles and doubles pattern fractions ratios are consistent with 1 within statistical errors. The EPIC-MOS cameras were being operated in the Small Window (SW) mode while the EPIC-pn camera was being operated in the Large Window (LW) mode. Background regions are chosen as circular regions with radii of 40 arcsec, close to the source region in the same unit on the camera. The background count rate remains $\approx 0.1\,\mathrm{cts\,s^{-1}}$. Then we create redistribution matrix files and auxiliary response files by running RMFGEN and ARFGEN tasks separately. We consider the EPIC spectra between 0.3–10 keV, unless otherwise specified. ADDSPEC is used to make a stacked spectrum for each camera, along with corresponding background spectra and response matrix files. The averaged EPIC-pn background count rate is $0.05205 \pm 0.00001\,\mathrm{cts\,s^{-1}}$ and the averaged EPIC-pn source count rate is $1.649 \pm 0.001\,\mathrm{cts\,s^{-1}}$.

### 2.2 *NuSTAR* Data Reduction

IRAS 13224−3809 was observed by the *NuSTAR* satellite seven times, listed in Table 2. The first observation interval (ObsIDs 60202001002, 60202001004 and 60202001006) was performed continuously followed by the second (ObsIDs 60202001008 and 602002001010) and the third intervals (ObsIDs 60202001012 and 60202001014), each slightly separated in order to better overlap with the *XMM-Newton* observations. We reduce the *NuSTAR* data using the standard pipeline NUPIPELINE V0.4.5, part of HEASOFT V6.19 package, and instrumental responses from *NuSTAR* caldb V20161021. Source spectra are selected from circular regions with radii of 35 arcsec, and the background is obtained from nearby circular regions with radii of 120 arcsec. Spectra are extracted from the cleaned event files using NUPRODUCTS for both FPMA and FPMB. The observed flux in the 3–30 keV band is also added in Table 2.

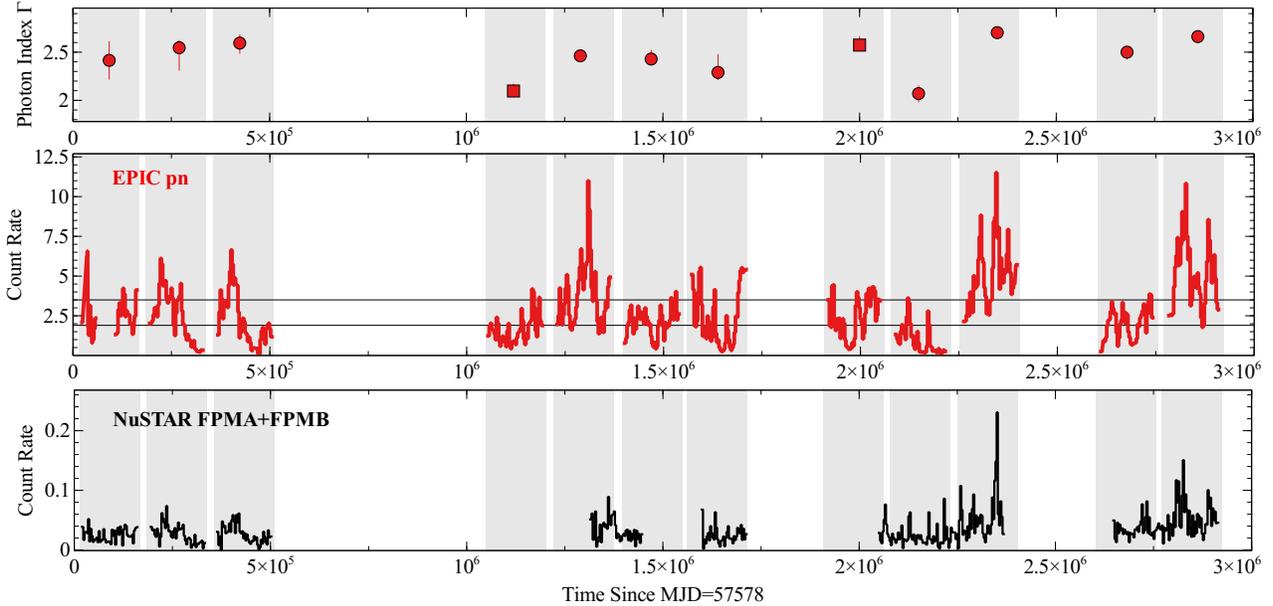

**Figure 1.** Top: the best-fit photon index of the powerlaw continuum Γ vs. time. (See Section 5 for more details). The circles are the best-fit values obtained by analyzing the *NuSTAR* FPM and *XMM-Newton* EPIC spectra simultaneously while the squares are the ones obtained by analyzing only the *XMM-Newton* EPIC spectra. Middle: the 0.3–10 keV EPIC pn light curve with 3 ks bins. The observing campaign is divided into 12 observation intervals, each of around 120 ks. The time coverage is shown in the grey shaded region. A time-resolved spectral analysis on each observation is introduced in Section 5. The light curve is also divided into three flux intervals, each with a similar number of counts, by horizontal solid lines. A flux-resolved spectral analysis of the three flux states is introduced in Section 4. Three peaks (12 cts s$^{-1}$) are detected during the whole observation. Bottom: the 3.0–78.0 keV *NuSTAR* FPMA+FPMB 3 ks light curve. The time axis has been coordinated with the *XMM-Newton* EPIC pn light curve. The *NuSTAR* observing campaign is separated into seven intervals. The 4th, 6th and 7th *NuSTAR* observations are divided into 2 time slices to conduct simultaneous time-resolved spectral analysis with the *XMM-Newton* data in Section 5.

### 2.3 Light Curves and Time Variability

The 0.3–10 keV EPIC pn light curve of 2016 observation is shown in the middle panel of Fig. 1. The data are grouped into bins with interval widths of 3 ks. Extreme flux peaks (12 ct s$^{-1}$) happened three times during the 1.5 Ms observing campaign with the count rate 3 times the average level (4 ct s$^{-1}$) and 100 times the lowest level (0.13 ct s$^{-1}$) The 0.3–10 keV band light curve shows rapid variability with a time scale of kiloseconds. The first flux peak happened in the gap between two *NuSTAR* observations, but the *NuSTAR* FPMs capture the second and the third flux peak in the *XMM-Newton* observation. This new observing campaign captures stronger flux peaks, compared with the light curves of the *XMM-Newton* observations in 2011 (e.g. 8 ct s$^{-1}$ in Chiang et al. 2015).

Fig. 2 shows the unfolded spectra of all the *XMM-Newton* and *NuSTAR* observations against a constant model. The 1–4 keV band shows 10 times flux difference between the highest (obs 0780561601) and the lowest flux state (obs 0792180301) while the iron band (4–7 keV) and the hard band (>10 keV) shows approximately 3 times flux difference and smaller variability than the 1–4 keV band. The averaged spectrum shows a strong soft component below 1 keV and a very broad strong Fe Kα emission line at the 4–7 keV band. The spectrum becomes softer and shows a weaker Fe Kα emission at higher flux states.

In order to group the spectra and reconstruct the response matrix in a more optimal way, we use the method in Kaastra & Bleeker (2016) by taking both every bin's averaged energy and photon counts into account[1]. The unfolded spectra against a constant model of all the observations show shown in Fig. 2. The flux at the iron band and the <1 keV band is less variable than that in the 1-4 keV band (approximately 10 times difference).

For the spectral analysis, we use the XSPEC(12.9.1k) software package (Arnaud 1996) to fit all the spectra discussed, and C-statistics (Cash 1979) is considered in this work, as required by the spectral binning method. The $\chi^2$ test is not used due to possible biased estimation of errors (Kaastra & Bleeker 2016). The Galactic column density towards IRAS 13224−3809 is fixed at the nominal value $5.3 \times 10^{20}$ cm$^{-2}$ from Kalberla et al. (2005) if not specified. The column density calculated by the method in Willingale et al. (2013) is $6.78 \times 10^{20}$ cm$^{-2}$ after accounting for the effect of molecular hydrogen. In the end of our stacked spectral analysis in Section 3.3, we will measure the Galactic column density with our X-ray spectra and fix it at our best-fit value in the following analysis. The photoionization cross section is from Balucinska-Church & McCammon (1992) and He cross section is from Yan et al. (2001). The solar abundances of Wilms et al. (2000) were used. The fit parameters are reported in the observed frame. The quoted errors of best fit parameter values are at the 90% confidence level. The best-fit model parameters are all reported in the observer's frame unless explained explicitly. We assume the cosmological parameters $H_0 = 73\,\mathrm{km\,s^{-1}\,Mpc^{-1}}$, $\Omega_{\mathrm{matter}} = 0.27$, and $\Omega_{\mathrm{vacuum}} = 0.73$. An isotropic model of emission is adopted when calculating the source luminosity. For local galactic absorption, the `tbnew` model (Wilms et al. 2000) is used. An additional constant model `constant` has been applied to vary normalizations between different instruments to account for calibration uncertainties.

---
[1] The python code is written by Carlo Ferrigino. https://cms.unige.ch/isdc/ferrigno/developed-code/

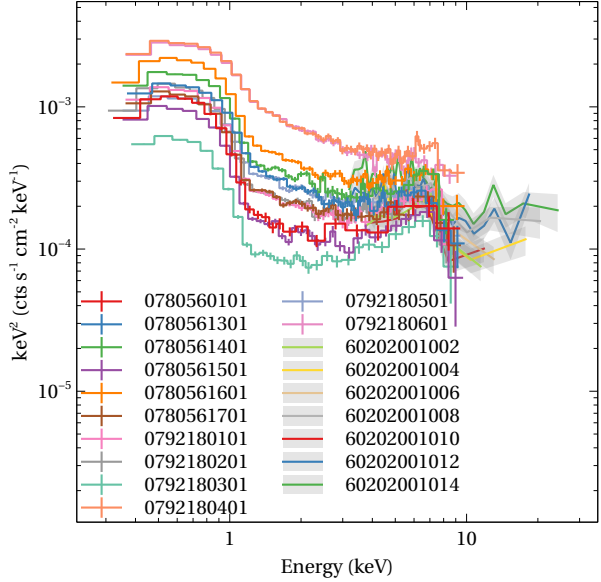

**Figure 2.** The unfolded spectra against a constant model of all the *XMM-Newton* (lines with errorbars) and *NuSTAR* (shaded lines) observations. The spectra have been grouped for clarity.

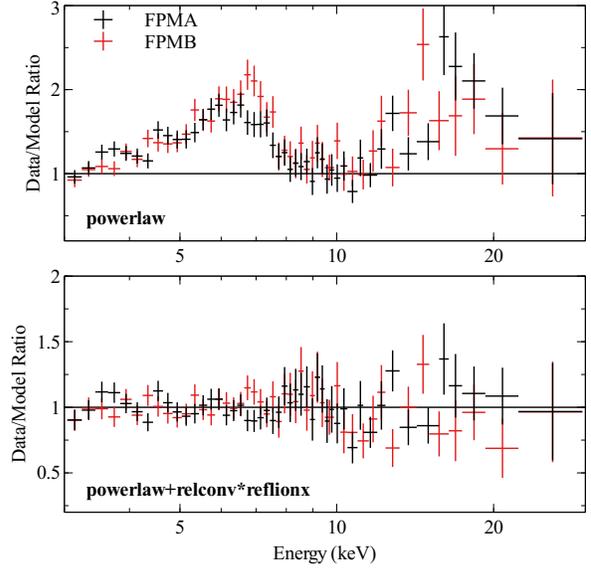

**Figure 3.** Top: the data/model ratio plot of two *NuSTAR* FPM spectra fitted with a simple Galactic absorbed powerlaw model. The ratio plot shows clear signatures of a disk reflection component, including a broad iron emission line and a reflection hump above 10 keV. Bottom: the ratio plot of only the FPM spectra fit with an absorbed power law plus a single relativistic disk reflection model. The total model reads `constant*tbnew*(powerlaw+relconv*reflionx)`. More details can be found in the first column of Table 3.

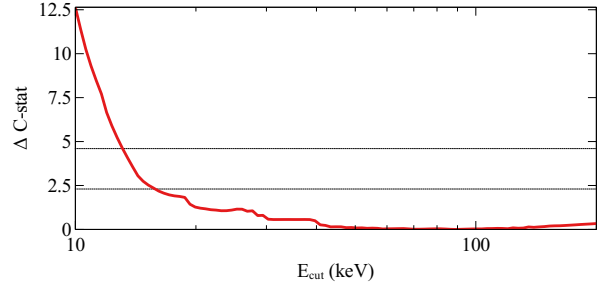

**Figure 4.** $\Delta C$-*stat* vs. powerlaw cutoff. The red solid line is obtained by fitting the 3-30 keV band of two *NuSTAR* FPM spectra. The two horizontal solid lines are $1\,\sigma$ and $2\,\sigma$ measurement lines.

Hereafter, black points with error bars are for FPMA; red points are for FPMB; green points are for MOS1; blue points are for MOS2; dark red points are for pn in figures unless explained specifically.

## 3 TIME-AVERAGED SPECTRAL ANALYSIS

In this section, we focus on studying the ionized reflection from the accretion disk around the central BH and identifying UFO absorption lines by analyzing the time-averaged spectra. We first fit *XMM-Newton* and *NuSTAR* data independently and then a quasi-simultaneous spectral analysis is conducted on both sets of data. The best fit model will be used as a template for further spectral analysis in Section 4 and Section 5.

### 3.1 Continuum Fit

#### 3.1.1 NuSTAR FPM Data

We first fit the stacked FPMA and FPMB spectra with an absorbed powerlaw. The ratio plot in the top panel of Fig. 3 shows a broad Fe line feature present around 6.4 keV and a Compton hump above 10 keV. To fit the reflection feature, an extended version of `reflionx` (Ross & Fabian

**Table 1.** XMM-Newton Observation Log. The duration is the length of scheduled observation. The 'annulus' column indicates where an annulus-shaped source region with an inner radius of 5 arcsec is used to extract spectra during the data reduction to reduce the pile-up effects. The usable percentage of each EPIC-pn exposure after correcting for flaring particle background is shown after the corresponding net exposure length. The observed flux is calculated by the best-fit model for EPIC pn spectra in the 0.3-10 keV band in the unit $10^{-12}$ ergs cm$^{-1}$ s$^{-1}$.

| Revolution | Obs.ID | Start Date | EPIC pn Net Exposure (ks) & Usage Percentage | Observed Flux ($10^{-12}$ ergs cm$^{-1}$ s$^{-1}$) | Annulus |
|---|---|---|---|---|---|
| 3037 | 0780560101 | 2016-07-08, 19:33:33 | 18 (95%), 36 (92%) | 3.68 ± 0.01 | MOS & pn |
| 3038 | 0780561301 | 2016-07-10, 19:25:31 | 122 (94%) | 3.77 ± 0.01 | MOS & pn |
| 3039 | 0780561401 | 2016-07-12, 19:34:13 | 78 (94%), 36 (95%) | 3.27 ± 0.01 | - |
| 3043 | 0780561501 | 2016-07-20, 19:01:53 | 120 (98%) | 2.57 ± 0.01 | - |
| 3044 | 0780561601 | 2016-07-22, 18:36:58 | 118 (98%) | 5.66 ± 0.01 | MOS & pn |
| 3045 | 0780561701 | 2016-07-24, 18:28:28 | 117 (97%) | 3.15 ± 0.01 | MOS |
| 3046 | 0792180101 | 2016-07-26, 18:18:44 | 123 (98%) | 3.36 ± 0.01 | MOS & pn |
| 3048 | 0792180201 | 2016-07-30, 18:02:21 | 120 (98%) | 3.52 ± 0.01 | MOS & pn |
| 3049 | 0792180301 | 2016-08-01, 17:54:51 | 108 (99%) | 1.53 ± 0.01 | - |
| 3050 | 0792180401 | 2016-08-03, 17:47:25 | 108 (94%) | 8.05 ± 0.01 | MOS & pn |
| 3052 | 0792180501 | 2016-08-07, 17:40:58 | 112 (97%) | 3.21 ± 0.01 | MOS |
| 3053 | 0792180601 | 2016-08-09, 18:29:52 | 116 (97%) | 7.84 ± 0.01 | MOS & pn |

**Table 2.** *NuSTAR* Observation Log. Similar with Table 1. The observed flux is the total flux of the best-fit model for FPMA and FPMB spectra in the 3-30 keV band in the unit of $10^{-13}$ ergs cm$^{-1}$ s$^{-1}$.

| Obs.ID | Start Date | Duration (ks) | Observed Flux ($10^{-13}$ ergs cm$^{-1}$ s$^{-1}$) |
|---|---|---|---|
| 60202001002 | 2016-07-08, 19:36:08 | 67.6 | 5.2 ± 0.1 |
| 60202001004 | 2016-07-10, 20:01:08 | 67.4 | 6.0 ± 0.3 |
| 60202001006 | 2016-07-12, 18:36:08 | 67.5 | 6.2 ± 0.3 |
| 60202001008 | 2016-07-23, 19:01:08 | 70.2 | 8.2 ± 0.4 |
| 60202001010 | 2016-07-27, 02:16:08 | 62.6 | 5.0 ± 0.3 |
| 60202001012 | 2016-08-01, 07:46:08 | 171.7 | 7.8 ± 0.2 |
| 60202001014 | 2016-08-08, 05:46:08 | 136.9 | 9.8 ± 0.3 |

2005) with wider iron abundance range (up to 30 times solar abundance) is used. The Fe K$\alpha$ line in `reflionx` is treated as the recombination lines of Fe XXV and Fe XXVI combined with the fluorescence lines of Fe VI-XVI. `relconv` is convolved with the local reflection model for relativistic effects (Dauser et al. 2013). After adding the relativistic reflection model `relconv*reflionx`, it reduces the statistics to $C-stat/\nu = 114.50/102$[2] (see the bottom panel of Fig. 3). The best-fit model parameters are listed in Table 3. The fit requires a central black hole with spin $a_* > 0.94$ viewed from an inclination angle of $i = 54$ deg. The fit puts a weak constraint on the iron abundance ($Z_{Fe} < 18$). In order to test a possible low energy cutoff in the broad band spectrum, the high energy cutoff parameter is allowed to vary and we obtained $E_{cut} > 15$ keV ($2\sigma$) by fitting the 3-30 keV band FPM spectra (refer to Fig. 4). Only the lower limit of the cutoff can be obtained. The curvature of the high energy emission can be well described with the Compton hump, part of the reflection model, with the high energy cutoff fixed at the maximum value in the following analysis. In Section 3.2.2, we will conduct a simultaneous 0.3–30.0 keV broad band analysis on both EPIC and FPM spectra.

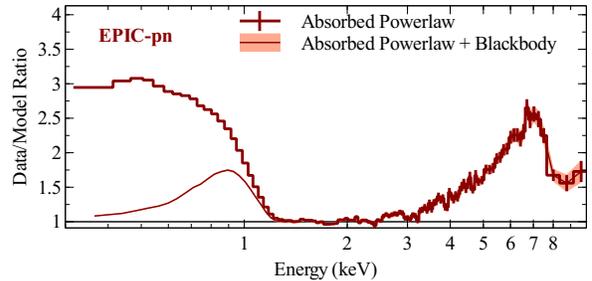

**Figure 5.** The data/model ratio plot for a fit of *XMM-Newton* EPIC-pn spectrum with a powerlaw model ($\Gamma = 2.8$), two simple relativistic lines and blackbody ($kT = 89$ eV). Shaded region: the two relativistic line models are removed to show the line shapes in the Fe K and L bands; points with error bars: two relativistic line models and `bbody` are all removed to show the soft excess below 1 keV. This ratio plot shows a strong soft excess, a strong disk relativistic Fe K and L emission line.

### 3.1.2 *XMM-Newton* EPIC Data

We first fit the 1–4 keV energy band of the EPIC spectrum with a simple absorbed powerlaw ($\Gamma = 2.8$). The ratio plot is extended to the whole band for only illustration purpose.

---

[2] $\nu$ is the number of degrees of freedom.

ble interpretation of the soft excess by a new high density disk reflection model. A distant reflection model fails to fit the broad Fe K emission line feature at the iron band (see the 1st panel of Fig. 6). So we initially use the simple and fast relativistic convolution model `kdblur2`, a convolution model adapted from `laor` (Laor 1991), which has a broken power law emissivity profile. For further analysis, we switch to the more sophisticated relativistic kernel model `relconv` (Dauser et al. 2013).

We first fit the soft excess with a soft cutoff powerlaw model (see the 2nd panel of Fig. 6) and a combination of a soft cutoff powerlaw and an additional relativistic reflection model (see the 3rd panel of Fig. 6). In the latter case, the photon index parameters in two `reflionx` models are tied to that of the `powerlaw` model. The iron abundance parameter is tied between two `reflionx` models. The additional reflection model decreases the residuals of the Fe L emission line but the cutoff model fails to fit the spectral shape below 0.7 keV. We fit the soft excess with a blackbody model `bbody` instead of a soft cutoff powerlaw, which decreases the statistics by more than 1000 (see the 4th panel of Fig. 6). In order to fit the residuals at <0.8 keV, we add another reflection model to fit the soft excess. It can reduce the residuals from $10\sigma$ to $4\sigma$ at energies below 0.6 keV and *C-stat* by 1000 (see the 5th panel of Fig. 6). Finally, we replace `kdblur2` with `relconv` for more accurate relativistic effects on the broad line features. A broken powerlaw emissivity profile is assumed. While the outer emissivity index is first fixed at 3 to meet the emissivity in flat spacetime, the inner emissivity index is left free to vary (see the second columns of Table 3). The inner edge of the accretion disk is assumed to be at the innermost stable circular orbit (ISCO) and the outer edge of the disk is fixed at $400\,R_g$ for simplicity. Limb-darkening effects are included in the model. The total continuum model combination now reads `constant*tbnew*(relconv*(reflionx1+reflionx2)+powerlaw+bbody)`. This model combination provides the best fit with $C\text{-}stat/\nu$=1237.73/454. The 6th panels of Fig. 6 and lower panel of Fig. 8 show the residual and data/model ratio plots respectively.

So far we have obtained the best-fit continuum model but there are still some narrow atomic features visible. For example, the absorption features in 1–2 keV, 2–4 keV, >8 keV band will be further discussed in Section 3.2.2 and be fitted with a photoionized absorber model `xstar`.

## 3.2 Detailed Spectral Analysis

In this section, in addition to the Fe absorption feature discussed in Parker et al. (2017b), we conduct a more detailed spectral analysis on *XMM-Newton* EPIC spectra first by identifying more UFO blueshifted absorption lines in the broad band fitting and then fitting these lines with the more physical ionized absorption model `xstar`.

### 3.2.1 Blueshifted Absorption Features

As discussed in the introduction section, a UFO from the disk of this source has been identified in this source (Parker et al. 2017b; Parker et al. 2017a). Here we present a detailed analysis and physical modeling of blueshifted Si XIV, S XVI,

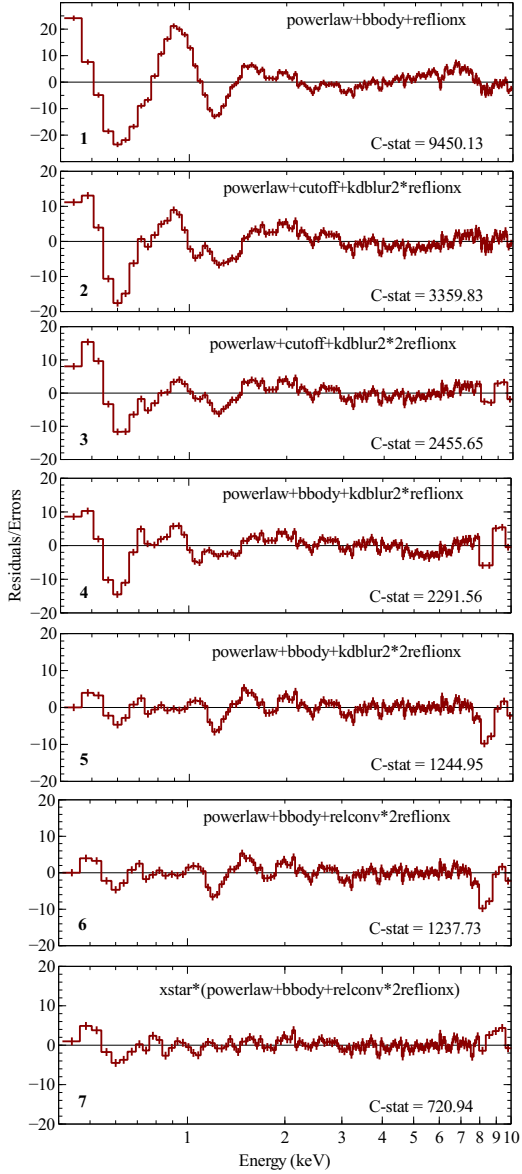

**Figure 6.** The residual/error plots for the different model combinations fit to the time-averaged *XMM-Newton* EPIC spectra (only the pn spectrum is shown here for clarity). All models are fit from 0.3–10.0 keV. The corresponding model and the statistics are marked in each panel. See the text for more details.

Fig. 5 shows the EPIC pn data/model ratio plot, indicating a strong soft excess below 1.5 keV. Two strong broad emission lines are shown in the Fe K and L bands, associated with the disk reflection and consistent with Ponti et al. (2010); Fabian et al. (2013); Chiang et al. (2015).

To fit the soft excess, we separately try a phenomenological single temperature blackbody model `bbody` and a cutoff powerlaw model `cutoffpl`. The normalization of the `bbody` model is defined as $L_{39}/D^2$, where $L_{39}$ is the source luminosity in $10^{39}$ erg s$^{-1}$ and $D_{10}$ is the distance in 10 kpc. In the end of this Section 3.4, we will discuss another possi-

**Table 3.** Best-fit Model Parameters for *NuSTAR* FPM (3.0–30.0 keV) and *XMM-Newton* EPIC (0.3–10.0keV) spectra. The flux is calculated by `cflux` in 0.3–10 keV band and in log base 10. The normalization of `bbody` model is defined as $L_{39}/D^2$, where $L_{39}$ is the source luminosity in $10^{39}$ erg s$^{-1}$ and $D_{10}$ is the distance in 10 kpc.

| Model | Parameter | Unit | Value | | | |
|---|---|---|---|---|---|---|
| | | | Only FPM relconv & reflionx | Only EPIC relconv & reflionx | FPM & EPIC relconv & reflionx | FPM & EPIC relxilld |
| `tbnew` | $N_{\rm H}$ | $10^{20}$cm$^{-2}$ | 5.3 | 5.3 | $6.39^{+0.22}_{-0.14}$ | $6.4 \pm 0.2$ |
| `xstar1` | $N_{\rm H1}$ | $10^{22}$cm$^{-2}$ | - | $14^{+34}_{-7}$ | $3.2^{+3.7}_{-0.3}$ | $7.2^{+0.3}_{-2.1}$ |
| | $\log \xi'_1$ | erg cm s$^{-1}$ | - | $3.64^{+0.22}_{-0.16}$ | $3.72^{+0.05}_{-0.12}$ | $3.62 \pm 0.03$ |
| | $Z'_{\rm Fe}$ | solar abundance | - | $3.1^{+0.5}_{-0.9}$ | $3.4 \pm 0.2$ | $2.0^{+1.3}_{-0.5}$ |
| | redshift $z_1$ | | - | $-0.187^{+0.004}_{-0.002}$ | $-0.189 \pm 0.003$ | $-0.188 \pm 0.002$ |
| `xstar2` | $N_{\rm H2}$ | $10^{22}$cm$^{-2}$ | - | $0.75^{+0.14}_{-0.17}$ | $0.78^{+0.14}_{-0.15}$ | $0.82^{+0.12}_{-0.21}$ |
| | $\log \xi'_2$ | erg cm s$^{-1}$ | - | $3.070^{+0.011}_{-0.035}$ | $3.05 \pm 0.02$ | $3.11^{+0.19}_{-0.12}$ |
| | redshift $z_2$ | | - | $-0.154 \pm 0.004$ | $-0.152 \pm 0.002$ | $-0.157 \pm 0.002$ |
| `bbody` | T | eV | - | $94.7^{+0.7}_{-0.7}$ | $91.7^{+0.5}_{-0.4}$ | $98.5^{+1.1}_{-1.6}$ |
| | Normalization | $10^{-5}$ | - | $7.56^{+0.08}_{-0.10}$ | $6.04^{+0.12}_{-0.80}$ | $5.7 \pm 0.3$ |
| `relconv` | Incl | degree | $54^{+3}_{-4}$ | $63.35^{+0.18}_{-0.26}$ | $67.0^{+2.3}_{-1.2}$ | $77^{+3}_{-7}$ |
| | $a_*$ | J/M$^2$ | >0.94 | $0.9960^{+0.0013}_{-0.0028}$ | $0.989^{+0.002}_{-0.003}$ | > 0.975 |
| | Inner Emissivity Index | | $5.0^{+1.2}_{-0.7}$ | $6.57^{+0.16}_{-0.14}$ | $6.04^{+0.12}_{-0.20}$ | $7.6^{+0.5}_{-1.5}$ |
| | Outer Emissivity Index | | 3 (fixed) | 3 (fixed) | $2.85^{+0.71}_{-0.10}$ | $3.5^{+0.2}_{-2.1}$ |
| | $R_{\rm break}$ | $R_g$ | $3.2^{+1.4}_{-0.5}$ | $4.4^{+3.4}_{-1.2}$ | < 5.8 | <5.4 |
| reflection 1 | $\log \xi_1$ | erg cm s$^{-1}$ | $2.4 \pm 0.3$ | $3.22 \pm 0.07$ | $3.13^{+0.07}_{-0.04}$ | $2.50 \pm 0.07$ |
| | $Z_{\rm Fe}$ | solar abundance | < 18 | $24.14^{+2.1}_{-1.4}$ | $24^{+3}_{-4}$ | $6.6^{+0.8}_{-2.1}$ |
| | Electron Density $\log(n_e)$ | cm$^{-3}$ | 15 | 15 | 15 | >18.7 |
| | log(Flux) | erg cm$^{-2}$ s$^{-1}$ | $-12.4^{+0.3}_{-0.2}$ | $-11.78^{+0.07}_{-0.10}$ | $-11.70 \pm 0.03$ | $-11.86^{+0.02}_{-0.03}$ |
| reflection 2 | $\log \xi_2$ | erg cm s$^{-1}$ | - | $1.68^{+0.02}_{-0.13}$ | $1.48^{+0.14}_{-0.10}$ | 0 (fixed) |
| | log(Flux) | erg cm$^{-2}$ s$^{-1}$ | - | $-12.60^{+0.04}_{-0.06}$ | $-12.55^{+0.04}_{-0.07}$ | $-12.14 \pm 0.02$ |
| `powerlaw` | $\Gamma_{\rm FPMA}$ | | $2.44^{+0.51}_{-0.12}$ | - | $2.37^{+0.05}_{-0.02}$ | $2.78 \pm 0.12$ |
| | $\Gamma_{\rm FPMB}$ | | $2.5^{+0.6}_{-0.2}$ | - | $2.41 \pm 0.06$ | $2.82^{+0.04}_{-0.12}$ |
| | $\Gamma_{\rm MOS1}$ | | - | $2.45^{+0.53}_{-0.10}$ | $2.495^{+0.02}_{-0.03}$ | $2.69 \pm 0.04$ |
| | $\Gamma_{\rm MOS2}$ | | - | $2.46^{+0.18}_{-0.04}$ | $2.51 \pm 0.03$ | $2.71^{+0.05}_{-0.03}$ |
| | $\Gamma_{\rm pn}$ | | - | $2.45^{+0.20}_{-0.02}$ | $2.49 \pm 0.03$ | $2.69 \pm 0.03$ |
| | log(Flux) | erg cm$^{-2}$ s$^{-1}$ | $-12.37^{+0.14}_{-0.37}$ | $-11.78^{+0.03}_{-0.04}$ | $-11.74 \pm 0.02$ | $-11.67 \pm 0.02$ |
| $C$-stat/$\nu$ | | | 114.50/102 | 720.94/448 | 797.73/567 | 822.32/566 |

**Table 4.** The best-fit absorption line parameters, including line energy in observer frame, FWHM, equivalent width and the $\Delta C$-stat value. $C$-stat/$\nu$ obtained by the continuum model in section 3.1 is 1237.73/455.

| Line | E (keV) | FWHM (keV) | EW (eV) | $\Delta C$-stat |
|---|---|---|---|---|
| Ne X | $1.181 \pm 0.010$ | $0.24^{+0.2}_{-0.05}$ | $37.5^{+0.2}_{-0.3}$ | 280.79 |
| S XVI | $3.16 \pm 0.03$ | $0.42^{+0.07}_{-0.05}$ | $50.90^{+11.0}_{-1.5}$ | 81.75 |
| Mg XII | $1.769^{+0.02}_{-0.03}$ | $0.12^{+0.07}_{-0.05}$ | $10.0^{+0.2}_{-0.4}$ | 54.6 |
| Si XIV | $2.42 \pm 0.03$ | $0.09 \pm 0.07$ | $10.1^{+4.0}_{-1.2}$ | 28.69 |
| Ar XVIII | 4.0 (fixed) | 0.001 fixed | <2.7 | - |
| Ca XX | 4.9 (fixed) | 0.001 fixed | <6.9 | - |

Ne X, Mg XII absorption features in the EPIC spectra (see the 6th panel of Fig. 6 for the residual plot and Fig. 8 for the ratio plot).

We first fit the absorption feature at around 1.2 keV in the observed frame with a Gaussian line model `gauss` and obtain a better fit with $C$-stat reduced by 280.79 for three additional degrees of freedom (line energy, FWHM and normalization). The best fit line energy of Ne X absorption line is $1.258^{+0.011}_{-0.009}$ keV in the source rest frame ($1.181 \pm 0.010$ keV in the observer frame) and consistent with the measurement in the RGS data ($10.0 \pm 0.5$Å, Parker et al. (2017b)) within the measurement error. Leighly et al. (1997) also found similar absorption features at 1–1.3 keV band in the ASCA data. Similarly to the Ne X line, we applied three more Gaussian line models for the rest of the absorption features. Parameters are given in Table 4. The blueshifted Ne X and S XVI are stronger and broader than the other two lines. In order to put limits on the strength of the Ar XVIII and Ca XX lines found in the PCA analysis in Parker et al. (2017a), we fit the Gaussian line models to EPIC spectra with the line energy parameter fixed as 4.0 keV and 4.9 keV. The inclusion of the two lines does not offer significant improvement to the fit and the equivalent width is negligible (see the last two rows of Table 4).

### 3.2.2 xstar *Modelling*

For further study, we model the absorption features in the 1–5 keV band (in Fig. 7) and Fe absorption feature above

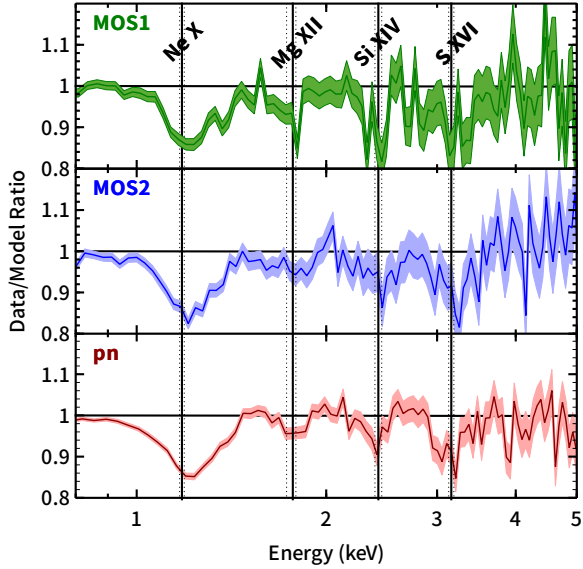

**Figure 7.** The data/model ratio plots for a time-averaged *XMM-Newton* EPIC spectral fit with the continuum model plus four additional Gaussian absorption lines. The Gaussian absorption line models in this plot are removed to show the line profiles. Blueshifted Ne X, Mg XII, Si XIV and S XVI absorption features are visible and labelled in 1–4 keV band. The best fit line parameter values can be found in Table 4. The vertical solid lines are the best-fit line energy and the dotted vertical lines are the measurement errors of the corresponding line energy values.

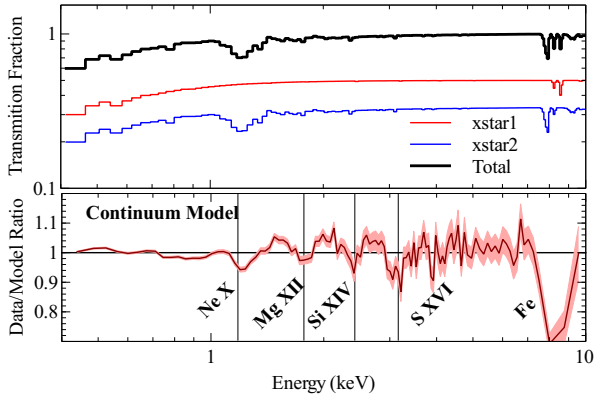

**Figure 8.** Top panel: Best-fit xstar transmission model convolved with the instrumental spectral resolution after being grouped (in black) and two different ionized components (the more ionized absorber xstar1 in red and the less ionized absorber xstar2 in blue). Bottom panels: The ratio plots for time-averaged *XMM-Newton* EPIC pn spectra fit without xstar (only pn is shown here for clarity). All the relevant absorption lines are labeled with dashed lines.

8 keV (in Fig. 5) with the physical model xstar and try to measure the overall ionization level and the averaged line-of-sight velocity of the disk outflow.

We construct custom photoionized plasma absorption models with xstar (Kallman & Bautista 2001). The grids are calculated assuming solar abundances except for that of iron, a fixed turbulent velocity of 2000 km s$^{-1}$ (Parker et al. 2017b) and an ionizing luminosity of 10$^{43}$ erg s$^{-1}$. Free parameters are the ionization of the plasma ($\log \xi'$) [3], the column density ($N_H$), the iron abundance ($Z_{Fe}$) and the redshift (z).

We first fit the EPIC spectra with only one xstar model and it reduces the *C-stat* by 99.23. See the blue line in the top panel of Fig. 8 for the model shape. This model perfectly fits the absorption features below 5 keV, including the blueshifted Ne X, S XVI, Mg XII and Si XIV absorption lines identified in Section 3.2.1. However, a single xstar model is not broad enough to fit the Fe absorption feature above 8 keV. A second xstar model is required to fit the blue wing of the Fe absorption line (see the red line in the top panel of Fig. 8 for the model shape). These two additional xstar models provide a significant improvement of fit by reducing the *C-stat*/ν of the EPIC spectra fit by 516.79 to 720.94/448. The iron abundances of the two xstar models are tied together. Parameters are given in Table 3. According to the best fit model in the top panel of Fig. 8, the more ionized absorber ($\log(\xi'_1/\mathrm{erg\,cm\,s^{-1}}) = 3.6$) is also more blueshifted than the less ionized absorber ($\log(\xi'_2/\mathrm{erg\,cm\,s^{-1}}) = 3.1$). The residual plot for the fit is shown in the bottom panel of Fig. 6. Two absorbers require column density $N_{H1} = 14^{+34}_{-7} \times 10^{22}$ cm$^{-2}$, $N_{H2} = 0.75^{+0.14}_{-0.17} \times 10^{22}$ cm$^{-2}$ and redshift $z_1 = -0.187^{+0.004}_{-0.002}$, $z_2 = -0.154 \pm 0.004$ respectively corresponding to line-of-sight outflowing velocity of v=0.26, 0.23c. In the following section, we will conduct simultaneous spectral analysis on both *NuSTAR* and *XMM-Newton* data using the same model combination.

### 3.3 Simultaneous Spectral Fitting

Finally, we present the simultaneous fit of *NuSTAR* FPM and *XMM-Newton* EPIC spectra based on the best-fit continuum model in Section 3.1 and the outflow model in Section 3.2.2. The total model is constant*tbnew*xstar*powerlaw+bbody+relconv*(2reflionx). The photon index parameters of different instruments are left to vary independently because this source is extremely variable and *NuSTAR* and *XMM-Newton* observations are not strictly simultaneous (compare two light curves in Fig. 1). We assume a broken powerlaw emissivity for the disk reflection component and allow both emissivity indices to vary.

The best-fit parameter values are listed in the third column of Table 3. The best-fit model is shown in Fig. 9. The two xstar models have blueshift up to $z_1 = 0.189 \pm 0.003$ and $z_2 = 0.152 \pm 0.002$, corresponding to velocity of $v_1 = 0.267^{+0.004}_{-0.003}$ c and $v_2 = 0.225 \pm 0.002$ c respectively for bulk motion in the radial direction. The best-fit column density of the outflow is $3.2^{+0.7}_{-0.2} \times 10^{22}$ cm$^{-2}$ and $0.78^{+0.14}_{-0.15} \times 10^{22}$ cm$^{-2}$, which is slightly lower than the value obtained by analyzing the RGS spectra ($9.5 \pm 0.5 \times 10^{22}\,\mathrm{cm}^{-2}$ Parker et al. 2017b).

---

[3] The prime symbol is to distinguish the ionization parameter of the outflow from that of the disk.

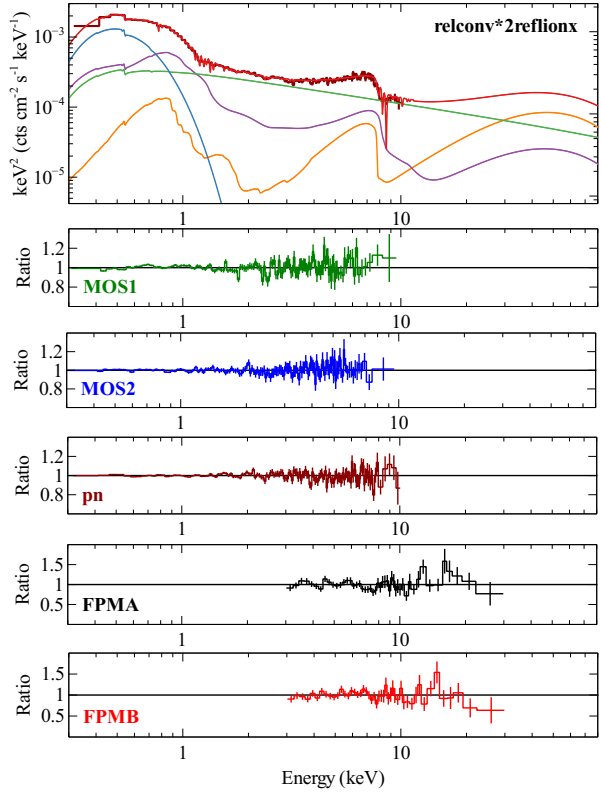

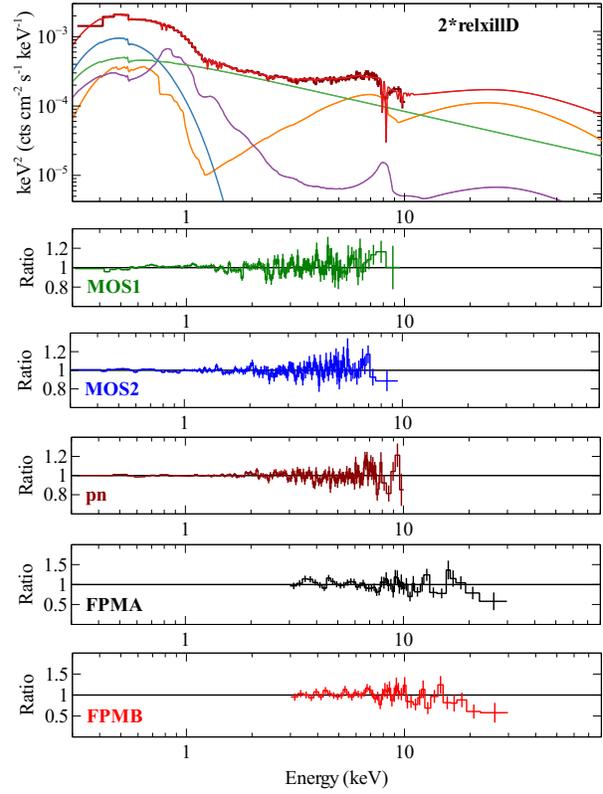

**Figure 9.** The best-fit EPIC pn model for the simultaneous fit of the quasi-simultaneous *XMM-Newton* and *NuSTAR* data. The rest-frame disk reflection model used here is `reflionx`. Green: powerlaw continuum; blue: blackbody model; purple: the disk reflection component with higher ionization; yellow: the disk reflection component with lower ionization; red: the total model. The dark red with errorbars are the unfolded spectra of EPIC pn. The bottom 5 panels show the ratio/model plots of *XMM-Newton* EPIC spectra and *NuSTAR* FPM fitted with the best-fit model.

**Figure 10.** Same as Fig. 9. The relativistic disk reflection model used here is `relxillD`.

The averaged spectrum in this analysis is harder than the previous fit with the same model combination in Chiang et al. (2015) ($\Gamma = 2.71 \pm 0.02$), which could be due to the change of the averaged temperature or optical depth of the corona through the whole observation. The best-fit relativistic parameters indicate a fast rotating black hole ($0.989^{+0.002}_{-0.003}$) viewed from an inclination of $67.0^{+2.5}_{-1.2}$ deg. The different emissivity index from the result in Chiang et al. (2015) may be due to a change of the geometry of the primary source (Wilkins et al. 2015). For example, the larger inner emissivity index means that the primary power law photons are more concentrated on the inner disk due to the light bending effects. A very steep emissivity profile indicates a very compact primary X-ray source.

The absorber has $Z_{\mathrm{Fe}} = 3.4 \pm 0.2$ solar iron abundance while the two disk reflection components require $Z_{\mathrm{Fe}} > 20$. This is not expected as the outflowing wind from the disk should share the element abundances of the disk. However, in Section 3.4 we find a good fit with a high density disk reflection model which only requires the disk iron abundance $Z_{\mathrm{Fe}} = 6.6^{+0.8}_{-2.1}$ so can potentially solve this problem.

The best-fit Galactic column density obtained by analyzing our X-ray spectra is $6.39^{+0.22}_{-0.14} \times 10^{20}\,\mathrm{cm}^{-2}$, which is slightly higher than $5.3 \times 10^{20}\,\mathrm{cm}^{-2}$ predicted by Kalberla et al. (2005) but consistent with the value obtained in Willingale et al. (2013) ($N_{\mathrm{H}} = 6.78 \times 10^{20}\,\mathrm{cm}^{-2}$). We will use our best-fit Galactic column density value in the following analysis.

### 3.4 High Electron Density Reflection Model

A very high iron abundance ($Z_{\mathrm{Fe}} = 24$) is required by fitting the soft excess with relativistic reflection model combination `relconv*reflionx` which assumes a constant electron density $n_e = 10^{15}\,\mathrm{cm}^{-3}$. In order to obtain a more reliable measurement on the disk iron abundance, we also try to fit the *XMM-Newton* and *NuSTAR* spectra with the recently developed model `relxillD` (García et al. 2016) which allows the electron density parameter to vary between $n_e = 10^{15}$ and $10^{19}\,\mathrm{cm}^{-3}$ while `reflionx` assumes $n_e = 10^{15}\,\mathrm{cm}^{-3}$. At higher $n_e$, the spectrum shows a higher temperature `bbody` shaped soft excess due to the increased influence of the free-free process on the spectrum at higher densities. Free-free absorption increasingly constrains low energy photons, increasing the temperature of the top layer of the disk and turning the reflected emissions at energies below 1 keV into a quasi-blackbody spectrum. García et al. (2016) show that a high electron density disk in 1H 0707−495 with $n_e \approx 10^{19}$ cm$^{-3}$ reduces the amplitude of the soft excess by 30%. Tomsick et al. (2018) obtain a better fit of the intermediate state spectra of Cyg X-1 with the electron density as a free parameter ($n_e = (3.98^{+0.12}_{-0.25} \times 10^{20}$ cm$^{-3}$) and only solar iron abundances.

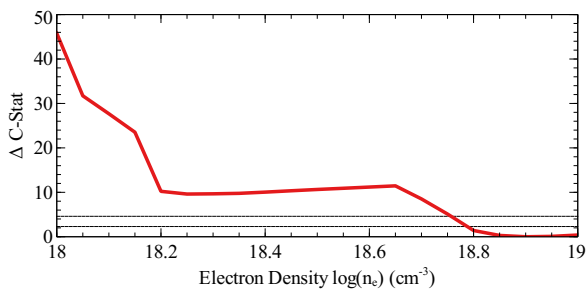

**Figure 11.** $\Delta C$-*stat* vs. electron density value in logarithmic units. The two horizontal solid lines are 1 $\sigma$ and 2 $\sigma$ measurement lines.

Here we fit the *XMM-Newton* EPIC and *NuSTAR* spectra in the 0.3–30 keV energy band with a `powerlaw`, a `bbody` and two relativistic reflection models `relxillD`. The reflection fraction parameters of the two `relxillD` are fixed as −1 to return only the reflection components. The cutoff energy parameter $E_{\rm cut}$ is fixed as 300 keV. The best fit model is shown in the top panel of Fig. 10 and it can offer a good fit with $C$-*stat*/$\nu$=822.32/566. One low-ionization ($\log(\xi_1/{\rm erg\,cm\,s^{-1}})$=0) and one ionized ($\log(\xi_2/{\rm erg\,cm\,s^{-1}})$=2.50±0.07) reflection component are required. The best fit iron abundance $Z_{\rm Fe} = 6.6^{+0.8}_{-2.1}$, which is much lower than the iron abundance obtained in Section 3. The two `rexillD` models require disk density $n_e > 10^{18.7}$cm$^{-3}$ (see Fig.11 for reference). The fit almost reaches the upper limit of the $n_e$ parameter in the current model. The best-fit high density disk reflection model is shown in Fig. 10. With such a high density, the soft band of the two reflection models has more emission than the best-fit `reflionx` models shown in Fig.9. This is still work in progress and further fits will be presented in a future paper (Jiang et al. in prep).

## 4 FLUX-RESOLVED SPECTRAL ANALYSIS

Parker et al. (2017b) found that the Fe, O and Ne absorption features are strongly flux dependent, suggesting that the UFO responds to the AGN continuum. In order to confirm whether the other absorption lines identified in Section 3.2.2 follow the same rule and probe more information on the overall spectral characteristics at different flux levels we conduct a broad band spectral analysis on three flux resolved spectra.

We divide the EPIC pn dataset into three flux levels, HF (high flux), MF (middle flux) and LF (low flux). The flux levels are chosen to have a similar number of total counts, as in Parker et al. (2017b).

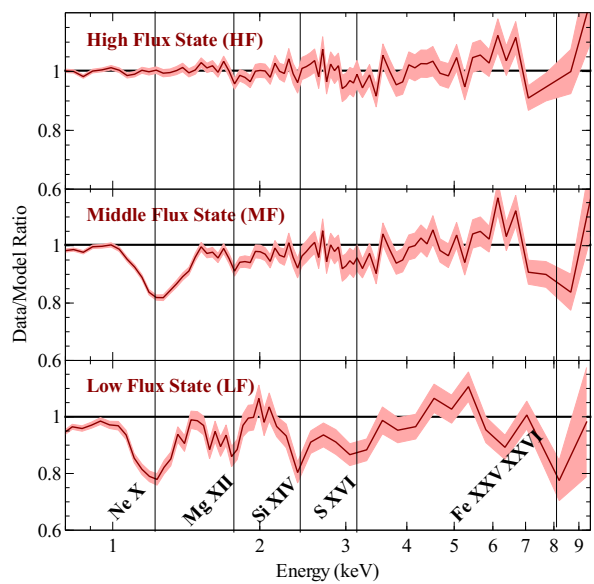

**Figure 12.** The ratio plot of the fits for three flux-resolved spectra, high flux (HF) state, middle flux (MF) state and low flux (LF) state to the best-fit UFO absorbed continuum model. The `xstar` models are removed in this plot to show the line shapes more clearly. All the relevant blueshifted absorption features are marked with solid black lines.

We first fit all three flux-resolved spectra with the same model template discussed above, `tbnew*xstar*(powerlaw+relconv*2reflionx+bbody)`. As the spin measured in the time-averaged analysis approaches to the spin limit in General Relativity, and is consistent with previous results (Ponti et al. 2010; Fabian et al. 2013; Chiang et al. 2015; Parker et al. 2017b), we fix it at the best-fit value obtained in the time-averaged analysis. The data/model ratio plots after removing the `xstar` absorption components to display the line profiles are shown in Fig. 12.

The powerlaw continuum and the corresponding disk reflection components show large different among three flux states. The `bbody` shows the lowest flux in the LF state, indicating a weakest soft excess. The 0.3-10 keV band flux values of all the components increase with increasing flux levels. The coronal emission shows a softer continuum ($\Gamma$=2.66 for HF state) compared with two lower flux states (e.g. $\Gamma$ = 2.03 for LF state). The reflection components and the primary powerlaw continuum show higher flux at higher flux state. The primary powerlaw emission shows more than 10 times flux difference between the LF and HF states while the flux of the reflection components only show 4 times difference. This can be explained by the light-bending model: the reflection component is less affected by the light bending effects as more lights are focused onto the inner area of the disk, resulting in increasing reflection fraction.

The best fit parameter values obtained by including the `xstar` models can be found in Table 5. Moreover, the blueshifted Mg XII, Si XIV and S XVI absorption signatures are only visible in the lower flux states, MF and LF. To fit these absorption features, the same `xstar` model generated in Section 3 is applied to LF and MF spectra. xstar models significantly reduce the residuals from the continuum fit. In order to obtain the upper limit of the absorption features in HF state, we fit the HF spectrum with one xstar model ($\log(\xi'/$erg cm s$^{-1}) \equiv 3.0$ and $z \equiv$ -0.13). The column density we obtain is $N_H < 2 \times 10^{20}$ cm$^{-2}$. The best-fit outflow column density for the LF state is higher than the best-fit value for the MF state, showing a flux-dependent outflow.

Table 5. Best-fit parameter values for three flux-resolved spectra. The units of the parameters are defined the same as in Table 3.

| Model | Parameter | High Flux (HF) | Middle Flux (MF) | Low Flux (LF) |
|---|---|---|---|---|
| xstar1 | $N_{H1}$ ($10^{22}$cm$^{-2}$) | - | < 1.0 | $4.2^{+1.3}_{-2.1}$ |
| | $\log \xi'_1$ (log(erg cm s$^{-1}$)) | - | $4.3^{+0.8}_{-0.7}$ | $3.7^{+0.2}_{-0.5}$ |
| | redshift $z_1$ | - | $-0.170^{+0.002}_{-0.005}$ | $-0.175 \pm 0.006$ |
| xstar2 | $N_{H2}$ ($10^{22}$cm$^{-2}$) | - | $0.6 \pm 0.4$ | $13.7 \pm 1.4$ |
| | $\log \xi'_2$ (log(erg cm s$^{-1}$)) | - | $3.07 \pm 0.07$ | $3.0^{+0.4}_{-0.3}$ |
| | redshift $z_2$ | - | $-0.143^{+0.05}_{-0.02}$ | $-0.132 \pm 0.007$ |
| bbody | kT (eV) | $93.7^{+0.2}_{-1.2}$ | $91.3^{+0.2}_{-1.1}$ | $92.3^{+0.3}_{-0.5}$ |
| | norm ($10^{-5}$) | $9.7 \pm 0.3$ | $8.15^{+0.12}_{-0.20}$ | $3.63^{+0.07}_{-0.11}$ |
| relconv | Inner Emissivity Index | $5.2^{+0.4}_{-0.3}$ | $6.3^{+1.2}_{-2.0}$ | $6.0^{+1.0}_{-3.5}$ |
| | R ($R_g$) | < 9 | < 8 | $69^{+11}_{-56}$ |
| reflionx1 | $\log \xi_1$ (erg cm s$^{-1}$) | $3.36^{+0.02}_{-0.04}$ | $3.24^{+0.19}_{-0.06}$ | $2.70^{+0.17}_{-0.12}$ |
| | log(Flux) (erg cm$^{-1}$ s$^{-1}$) | $-11.41 \pm 0.03$ | $-11.84^{+0.06}_{-0.09}$ | $-12.58^{+0.08}_{-0.04}$ |
| reflionx2 | $\log \xi_2$ (erg cm s$^{-1}$) | $1.53^{+0.13}_{-17}$ | $2.12^{+0.04}_{-0.08}$ | $1.80^{+0.14}_{-0.21}$ |
| | log(Flux) (erg cm$^{-1}$ s$^{-1}$) | $-12.35^{+0.12}_{-0.15}$ | $-12.45^{+0.20}_{-0.12}$ | $-12.77 \pm 0.08$ |
| powerlaw | $\Gamma_{pn}$ | $2.66 \pm 0.04$ | $2.31^{+0.04}_{-0.02}$ | $2.03 \pm 0.04$ |
| | log(Flux) (erg cm$^{-1}$ s$^{-1}$) | $-11.33^{+0.03}_{-0.04}$ | $-11.90^{+0.07}_{-0.04}$ | $-12.28^{+0.04}_{-0.03}$ |
| | $C$-stat/$\nu$ | 166.79/135 | 157.46/120 | 141.58/119 |

## 5 TIME-RESOLVED SPECTRAL ANALYSIS

Fig. 1 shows the extreme variability of IRAS 13224-3809 in the EPIC 0.3–10 keV band on timescales of ks or less. In order to further study the spectral variability with respect to time and the averaged flux, we conduct spectral analysis on each of the twelve observations. A simultaneous 0.3–30 keV broad band spectral analysis is conducted if simultaneous *NuSTAR* data are available during that *XMM-Newton* observation. The simultaneous *NuSTAR* spectra are extracted according to the time coverage of the *XMM-Newton* observations (compare the grey shaded region and *NuSTAR* combined FPM light curve Fig. 1).

All the time-resolved spectra are fitted with the same model combination as the one obtained above, with relativistic parameters and the column density fixed at the best-fit value obtained in the time-averaged analysis. RDC and PLC are, respectively, the reflection and powerlaw continuum flux between 0.3–10 keV in log scale in the unit of erg cm$^{-2}$ s$^{-1}$ calculated by cflux model in XSPEC. The reflection fraction is simply defined as the ratio of the reflection flux over the power-law continuum flux. The bbody normalization is defined as $L_{39}/D_{10}^2$, where $L_{39}$ is the source luminosity in units of $10^{39}$erg s$^{-1}$ and $D_{10}$ is in unit of 10 kpc. All the best-fit parameters are plotted in Fig. 13.

The following conclusions can be drawn from Fig. 13:

(i) When the blackbody flux is higher, the temperature of the blackbody tends to be higher, indicating a steeper soft excess. The temperature and luminosity of the bbody component follows the Stefan-Boltzmann $F \propto T^4$ relation, as found in the previous study of Chiang et al. (2015), indicating a constant emission area of the soft excess. The flux of the blackbody follows the same trend as that of the powerlaw component (see the first panel of Fig. 13).

(ii) The reflection component is highly positively correlated with the powerlaw continuum. However, as the reflection component is less variable than the powerlaw component, the reflection fraction is negatively correlated with the powerlaw flux (refer to the second and the third panel of Fig. 13). They show that the flux of the reflection components in the 0.3–10.0 keV change by a factor of 3-4 and the flux of the powerlaw component changes by a factor of 30. This can potentially be explained by the light-bending model: when the corona is closer to the black hole, the powerlaw flux decreases drastically due to more photons lost to the event horizon. The reflection component is less affected by the light bending affects as more light is focused onto the inner disk, resulting in an increasing reflection fraction.

(iii) From the Spearman's Rank Correlation (SRCC = −0.634, P − value = 0.27 using t-distribution), we find that the inner emissivity index and the powerlaw flux have a weak negative correlation. This can also be explained by a change of primary source height in the lamp-post geometry as in the point above. The inner emissivity tends to be higher when the source is close to the black hole and the primary photons are more concentrated to the inner area of the disk due to the light-bending effects. The primary powerlaw component is weaker, caused by the loss of the primary photons to the central black hole (refer to the forth panel of Fig. 13).

(iv) In Fig. 1 we show the best-fit powerlaw photon index to compare with the light curve. There is a positive correlation between the source brightness in 0.3-10.0 keV band and the powerlaw photon index. The brighter the source is, the softer the powerlaw is.

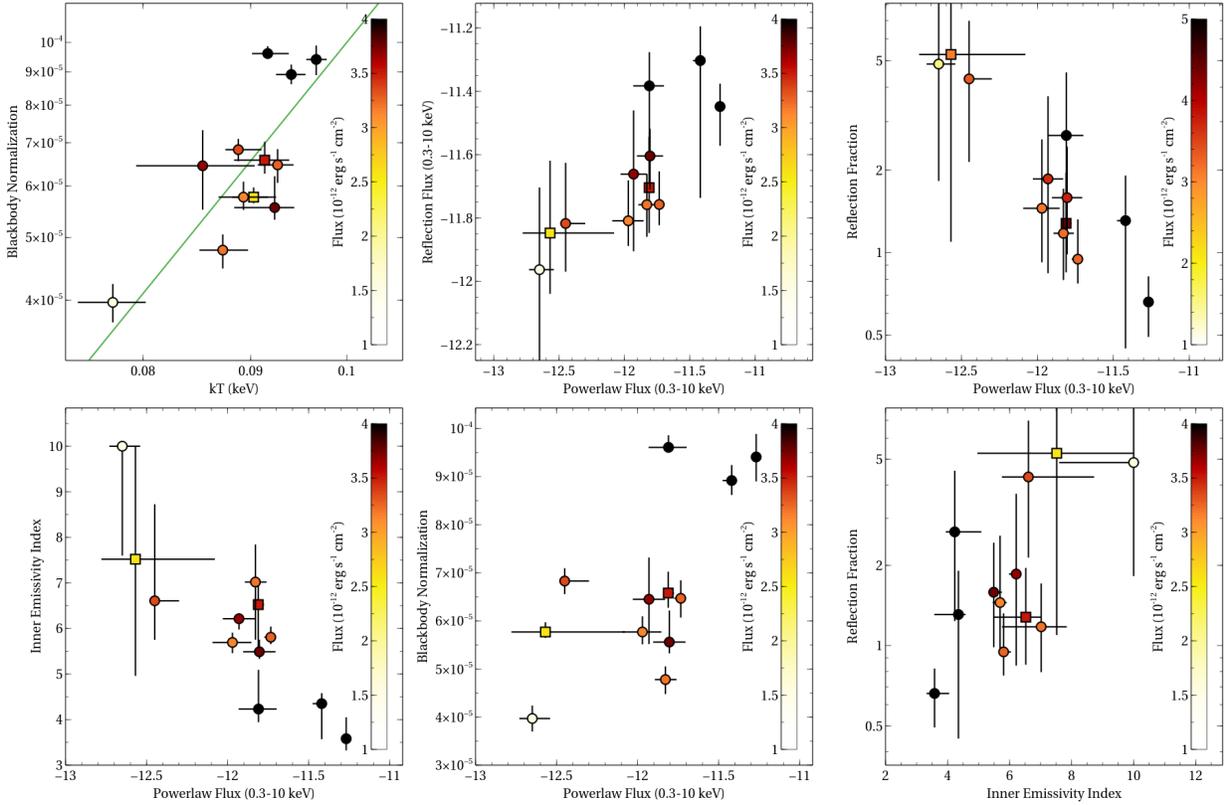

**Figure 13.** Time-resolved spectral analysis results. PLC and RDC are, respectively, the powerlaw continuum and the reflection flux in log scale calculated by `cflux` between 0.3–10 keV. The reflection fraction is defined as the ratio of the reflection flux to the power law continuum flux in that band. The Stefan-Boltzmann relation $F \propto T^4$ is plotted with a green line in the first panel for reference. The emissivity index here is the inner emissivity index while the outer emissivity index is fixed at 3. The color bar of the plot symbol indicates the average flux in the corresponding time slice. The best-fit values obtained by analyzing two *XMM-Newton* observations without simultaneous *NuSTAR* data are marked in squares.

## 6 DISCUSSION

### 6.1 Overall Bolometric Luminosity Estimation

In this section, we calculate the bolometric luminosity in three ways. We first apply a simple bolometric correction to the observed flux in the 2–10 keV band. Second, we apply the linear relation between the spectral index and the Eddington ratio found in Brightman et al. (2013). Thirdly, a continuum fitting method is applied to the UV/optical-X-ray band first with a phenomenological model (`bbody+powerlaw`). As the mass is uncertain, we leave a factor of $10^7 \ M_\odot/M$ in our estimates of Eddington fraction.

First, the best fit model for time-averaged spectra in Table 3 shows the source has flux of $6.87 \times 10^{-13}$ erg cm$^{-2}$ s$^{-1}$ in the 2–10 keV band. Given the measurement that the luminosity distance of IRAS 13224−3809 is 288Mpc, $L_{2-10\text{keV}} = 6.82 \times 10^{42}$ erg s$^{-1}$. $L_{\text{Edd}}$ for central object with mass $10^7 M_\odot$ is $12.6 \times 10^{44}$ erg s$^{-1}$. According to the estimation of the bolometric correction given for the 2–10 keV flux for NLS1 (see the top panel of Fig. 12 in Vasudevan & Fabian 2007), the overall bolometric luminosity $L^{\text{ave}}_{\text{bol}} = \kappa_{2-10\text{keV}} \times L_{2-10\text{keV}} \approx 50 \times 6.82 \times 10^{42}$ erg s$^{-1}$ = $3.4 \times 10^{44}$ erg s$^{-1}$. The best fit model flux for the flux peak spectrum is around $9.28 \times 10^{-13}$ erg cm$^{-2}$ s$^{-1}$ between 2–10 keV, corresponding to bolometric luminosity $L^{\text{peak}}_{\text{bol}} = 4.65 \times 10^{44}$ erg s$^{-1}$. This means that the source remains sub-Eddington both on average ($0.27 \frac{10^7 \ M_\odot}{M} L_{\text{Edd}}$) and at peaks ($0.37 \frac{10^7 \ M_\odot}{M} L_{\text{Edd}}$) according to the calculation of only the X-ray band. This result is roughly consistent with Sani et al. (2010).

Second, according to the estimation of the bolometric ratio given by the powerlaw photon index, $\lambda_{\text{Edd}} = 69.78$ for the peak spectra ($\Gamma = 2.86$) and $\lambda_{\text{Edd}} = 4.22$ on average ($\Gamma = 2.47$) by using the correlation $\Gamma = (0.32 \pm 0.05) \log \lambda_{\text{Edd}} + (2.27 \pm 0.06)$ (Brightman et al. 2013). The extremely high Eddington ratio obtained for the flux peak however is very uncertain, because the samples in Brightman et al. (2013) do not have sources with the primary continuum softer than $\Gamma = 2.2$.

Third, we try to estimate the accretion rate by fitting the SED with phenomenological model. The Optical Monitor (OM) on *XMM-Newton* was operated with only one filter during the observing campaign. So we extract the photometry data from simultaneous *Swift* (uvot) observations (Buisson et al. 2018). We used a circular source region of 5 arcsec radius and circular background region of 15 arcsec radius from a nearby source free area. UV fluxes have been corrected for Galactic reddening $E(B-V) = 0.06$. We first try to fit with the phenomenological continuum model `bbody+powerlaw`. The `bbody` model is for the disk thermal component and the `powerlaw` is for the primary continuum in the harder band. We fix the photon index of the `powerlaw` as 2.5, the same as we obtained in the time-averaged spectral analysis. The temperature of the thermal component is fixed at 1 eV and 10 eV to obtain upper limit and lower limits on the broad band luminosity. The fit with $kT$=1 eV gives a total model luminosity between 1-$10^4$ eV around $4 \times 10^{-11}$ erg cm$^{-2}$ s$^{-1}$, corresponding to $0.3 \frac{10^7 \, M_\odot}{M} L_{\rm Edd}$, and the fit with $kT$=10 eV gives $4 \times 10^{-10}$ erg cm$^{-2}$ s$^{-1}$, corresponding to $3.0 \frac{10^7 \, M_\odot}{M} L_{\rm Edd}$. Therefore, the estimation of the bolometric luminosity by the continuum fitting with a phenomenological continuum model is $0.3 - 3 \frac{10^7 \, M_\odot}{M} L_{\rm Edd}$. This measurement is consistent with the results obtained by Ponti et al. (2010), where they fit the UV and 2–10 keV *XMM-Newton* data with `diskpn` and estimated $\lambda_{\rm Edd} \approx 1$ for a $10^7$ $M_\odot$.

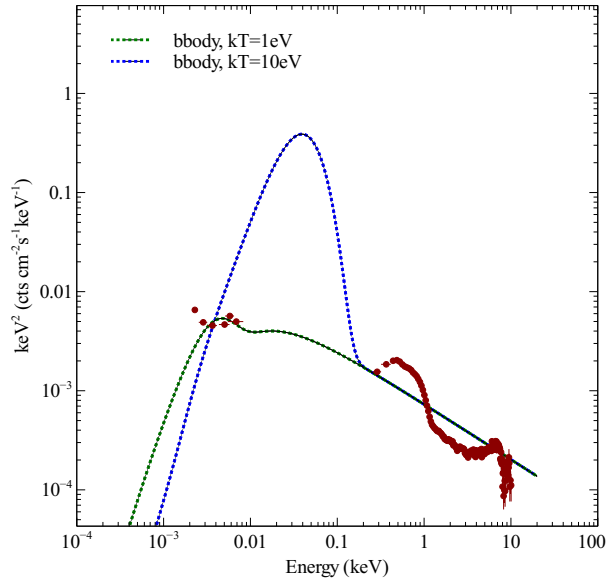

**Figure 14.** Continuum fitting with `bbody+powerlaw` to IRAS 13224−3809 *XMM-Newton* and *Swift* (uvot) data. The blue and green dotted lines are the best-fit phenomenological continuum models with $kT$=1, 10 eV. See the text for more details.

The estimation of the bolometric luminosity suggests that the source is close to the Eddington limit and may suffer accretion instability causing extreme X-ray variability. This may also be relevant to the launching of the UFO and the rapid X-ray variability (e.g. Done & Jin 2016). It is generally thought that the most powerful outflows are launched near to the Eddington limit, and powerful outflows are observed in the ultra-luminous X-ray sources (Pinto et al. 2016; Walton et al. 2016; Pinto et al. 2017b; Kosec et al. 2018) , at least some of which are super-Eddington neutron stars (Bachetti et al. 2014; Fürst et al. 2016; Israel et al. 2017a,b). Similarly, Leighly (2001) also found there are more high ionization emission lines observed in the UV band that are dom-inated by wind emission in the objects close the Eddington limit.

### 6.2 Soft Excess

In our analysis, the soft excess can be fit very well with a simple `bbody` component and an additional relativistic reflection component. The `bbody` temperature and flux follow $F \propto T^4$ relation, which indicate a constant emission area for the soft excess. Chiang et al. (2015) discussed possible explanations of the soft excess as reprocessing of both the coronal emission and the reflection. Strong gravitational effects can cause the thermal and reflected disk photons to return to the disk surface (Cunningham 1976). The innermost area of the disk is thus heated sufficiently to emit blackbody radiation in the soft band.

In Section 3.4, we explore the possibility of the soft excess as part of the disk reflection. `relxillD` is an extended version of `relxill` and models a relativistic reflection spectrum from an accretion disk with the electron density on the surface of the disk allowed to vary freely (`reflionx` assumes $n_{\rm e} \equiv 10^{15}$ cm$^{-3}$). We successfully fit the EPIC spectra in the 0.3–10 keV energy band by replacing `reflionx` with `relxillD` models. The fit favours `reflionx` over `relxillD` with smaller value of *C-stat*. But the best fit iron abundance $Z_{\rm Fe} = 6$, which is much lower than the iron abundance obtained in Section 3. We have obtained a strong constrain on the lower limit of the electron density $n_{\rm e} > 10^{18.7}$cm$^{-3}$ of the top layer of the disk by fitting the soft excess with a combination of high density reflection models and a phenomenological blackbody-shaped model. The two `xstar` models are consistent with the ones obtained in Section 3.2.2.

The high density disk reflection model proposed in García et al. (2016) is based on an extended model of the standard accretion disk. The electron density at high accretion rates is $n_e \dot{m}^2 \propto (1-f)^{-3}$, where $f$ is the fraction of power released by the disk onto the corona and $\dot{m}$ is the accretion rate in Eddington unit (Svensson & Zdziarski 1994). If 90 per cent of the disk power is taken by the corona (f=0.9), $n_e \dot{m}^2 \approx 10^{19} - 10^{18}$cm$^{-3}$ for black hole masses of $10^6$-$10^7$ $M_\odot$ (see Fig 1. of García et al. (2016)). At the electron density $n_e$ as high as $10^{18.7}$cm$^{-3}$, the reflection spectrum shows a higher temperature `bbody` shaped soft excess. This is due to the increased influence of the free-free process on the spectrum at higher densities. Free-free absorption increasingly constrains low energy photons, forcing the surface temperature of the reflecting material to rise and turning the reflected emissions at energies below 1 keV into a quasi-blackbody spectrum. The continuum shape can also depend on the density of the disk. The powerlaw continuum is usually interpreted as the thermal Comptonization of the seed photons from the disk. The spectral index of this continuum has a positive correlation with the corona temperature and the optical depth $\tau$ (Zdziarski et al. 1996). In our time-resolved spectral analysis result, the power law continuum is softer when the `bbody` temperature is higher. The change of powerlaw hardness is caused by either by a change of the temperature of the primary source or a change of geometry. Since the former factor increases when $n_e$ increases, $\tau$, the geometry dependent parameter, must change and the optical depth must be lower when the source is softer and brighter.

The current version of `relxillD` still requires improve-ment. For instance, more element abundances are expected to be higher than the solar level as well as iron. This is still work in progress and further fits will be presented in a future paper (Jiang et al. in prep).

### 6.3 Disk Reflection

To better model a turbulent accretion disk, two reflection components with different ionizations are used, one with a moderate ionization of $\log(\xi_1/\mathrm{erg\,cm\,s^{-1}})=3.13$ and one with a low ionization $\log(\xi_2/\mathrm{erg\,cm\,s^{-1}})=1.48$. The current model requires overabundant iron while the other elements are assumed to be solar due to the limitations of the `reflionx` model we use for analysis. However, as discussed above the need for a very high iron abundance and additional `bbody` can be potentially reduced by replacing the `reflionx` model with extended version of `relxillD` with higher disk density (discussed in Section 6.2). The disk iron abundance is $Z_\mathrm{Fe} = 6.6^{+0.8}_{-2.1}$ with `relxillD` and more consistent with the iron abundance of the UFO we obtain by fitting the absorption lines with `xstar`. The high iron abundance, even when fitted with high density disk reflection model, indicates that the other elements might be more abundant than solar. Wang et al. (2012) presented a strong correlation between the outflow strength in quasars, measured by the blueshift and asymmetry index (BAI), and the metallicity, measured by Si IV O IV/ C IV, based on the quasar samples built in the Sloan Digital Sky Survey. For example, a significantly higher metallicity ($Z > 5$) is indicated for quasars with BAI>0.7. The metallicity may play an important role and be connected with the quasar outflow.

Various authors (Martocchia & Matt 1996; Martocchia et al. 2000; Wilkins & Fabian 2012; Dauser et al. 2013) have shown that for simple coronal geometries the radial emissivity index decreases very sharply with source height in the most inner area of the disk (as low as $q = 1$ when the source height is up to $100\,R_g$). It tends to approximate as $q = 3$ in the outer area of the disk where the spacetime can be approximated to be flat. This effect reduces sharply with the source height. When the source height is $3\,R_g$, the emissivity profile is well approximated by a broken power law (outer emissivity index $q = 3$) with a very low break radius. A high inner emissivity index and a low break radius as in IRAS 13224−3809 indicates a very small source height ($<2\,R_g$ for instance).

IRAS 13224−3809 was observed by *XMM-Newton* for 500ks in 2011. Chiang et al. (2015) estimated the spectral variability during the observation and obtained a steeper emissivity profile for the time-averaged spectra (inner emissivity index $q_1 > 9$, outer emissivity index $q_2 = 3.4^{+0.3}_{-0.2}$ and $R_\mathrm{break} = 2.1 \pm 0.1\,R_g$) than we find. This could be caused by differences in the averaged source height in two observations. Moreover, the new campaign captures stronger flux peaks (12 cts/s) than Chiang et al. (2015) (8 cts/s), which is consistent with a higher source and therefore less extreme emissivity profile.

The second and the third panels of Fig. 13 show that the powerlaw continuum is more variable than the reflection component. The reflection fraction has an inverse relation with the powerlaw flux though the reflection component flux follows the same trend with the powerlaw flux. The light-bending solution (Miniutti et al. 2003) is poten- tially a good explanation for this: when the corona is closer to the central black hole the trajectories of more photons will be bent towards the black hole and more primary continuum photons will be lost. The reflection component is however less affected by the light bending effects, as more light is focused onto the inner disk, resulting in an increasing reflection fraction. Similar results have been found in other sources, such as MCG−6−30−15(Miniutti et al. 2003; Vaughan & Fabian 2004), NGC 3783 (Reis et al. 2012), and XRB XTE J1650−500 (Rossi et al. 2005; Reis et al. 2013), where the variability is dominated by the powerlaw continuum. An extreme case is the *NuSTAR* observation on Mrk 335 in 2013. Parker et al. (2014) found the reflection fraction decreases sharply with the increasing flux. The low-flux spectra of Mrk 335 are well described by only disk reflection model and indicates extreme light bending effects happening within $2\mathrm{R}_g$. The anti-correlation between the inner emissivity index and the flux of the power law continuum (the fourth panel of Fig. 13) in IRAS 13224−3809 also supports this interpretation. When the source is closer to the central black hole, the flux of the primary continuum decreases due to stronger light bending and the emissivity index is higher due to photons being focused onto the inner disk.

### 6.4 Ultra Fast Outflow

The combined spectral analysis of the stacked *NuSTAR* and *XMM-Newton* spectra shows two relativistic outflowing absorbers in the source. The more ionized absorber ($\log(\xi'_1/\mathrm{erg\,cm\,s^{-1}})=3.72^{+0.05}_{-0.12}$) has a higher blueshift ($z_1 = 0.18 \pm 0.003$, corresponding to line of sight velocity $v_1 = 0.267^{+0.04}_{-0.03}$ c) while the less ionized absorber ($\log(\xi'_2/\mathrm{erg\,cm\,s^{-1}})=3.05 \pm 0.02$) has a lower blueshift ($z_2 = 0.152 \pm 0.002$, corresponding to line of sight velocity $v_2 = 0.225 \pm 0.002$ c). We note that a similar UFO is found in 1H0707−495: Dauser et al. (2012) identified blueshifted narrow features at 2–5 keV band from H-like ions (Si, S, Ca) in 1H0707−495 *XMM-Newton* spectra as ultra-fast wind absorption features. The ionization of the disk wind in 1H0707-495 has small fluctuations ($\log(\xi/\mathrm{erg\,cm\,s^{-1}})\approx 3.5$) but the velocity has a difference of $0.07\,\mathrm{c}$ between different observations. Hagino et al (2016) also found evidence for an Fe absorption feature at 7.1–7.5 keV, with a velocity of 0.18c, also consistent with the velocity found by Dauser et al. (2012) (v=0.11–0.18c).

As found in Parker et al. (2017b), the inclusion of the UFO absorption lines does not have any significant impact on the measured reflection parameters. Similarly, in 1H0707−495, Dauser et al. (2012) find no large difference on the relativistic parameters, such as the corona height in the lamp-post scenario, the black hole spin and the viewing angle, after including the wind component. This is contrary to the results of Hagino et al. (2016), who found that the relativistic blurring parameters for 1H 0707-495 were less extreme when UFO absorption was taken into account. This difference is likely due to the much higher data quality and broader energy band used here, where we consider stacked spectra with very high signal-to-noise, whereas Hagino et al. (2016) examine individual observations and largely limit their analysis to the 2–10 keV band.

The lower ionization absorber is mainly used to fit the low energy absorption lines below 5 keV such as Ne X line and the red wing of the Fe absorption line, while the more ionized absorber is to fit the blue wing of the iron absorption above 8 keV. Parker et al. (2017b); Pinto et al. (2017a) fit the UFO absorption in the RGS and high-energy EPIC-pn spectra with a single absorber, with a velocity intermediate between the two we find here. The significant improvement that we find for fitting with two zones instead of one is likely an indication that there is structure or stratification of the UFO material, which we discuss below. The increasing level of ionization increases the rest-frame energy of the atomic feature lines and removes some of them as well (compare the red and blue lines in Fig. 8 for reference). Thanks to the high signal/noise of the *XMM-Newton* EPIC cameras in the soft band, in addition to the iron absorption feature at 8.1 keV found in the EPIC-pn spectrum (Parker et al. 2017b) we have identified Ne X (equivalent width 37.5 eV), S XVI (equivalent width 50.90 eV), Mg XII (equivalent width 10.0 eV) and Si XIV (equivalent width 10.1 eV) absorption lines (see Fig. 7 for reference). When the source is at high fluxes, the absorption features are weaker, consistent with the wind being more photoionised due to more photons emitted from the continuum source to the outflowing wind (see Fig. 12).

There are several different scenarios that could explain the observed outflow properties. An accelerating wind model was proposed in Murray et al. (1995) where the outflowing velocity increases along the outflow stream line. However, our data show that the faster absorber has higher photon ionization, which is contrary to the expectation in this model. Another possible interpretation is that the highly ionized layer lies inside a low ionized layer of the wind, and is more exposed to the continuum photons. This model was also used to explain the properties of the UV data of NGC 5548 in (Elvis 2000). Alternatively, Gallo & Fabian (2011) proposed that the absorption features result from ionized materials corotating with the disk in a surface layer. A different line of sight through this layer, caused by changes in coronal geometry, can result in a different observed optical depth and therefore different absorption feature. Such a model has been successfully applied to PG 1211+143 (Gallo & Fabian 2013). According to this model, the coexistence of a more ionized faster absorber and a less ionized slower absorber in our analysis can be interpreted as a layer in the inner disk which is more photoionized and faster than the layer further out.

These two absorbers do not show significant changes with time during the observation. It indicates a relatively constant outflow from the disk in the timescale of kiloseconds, confirming the result of Parker et al. (2017b); Pinto et al. (2017a), where we show that the Fe XXV/XXVI absorption feature has been approximately constant since 2011. This is interesting, as the UFO in PDS 456 , which has a much larger $M_{\rm BH} \approx 10^9 \, {\rm M_\odot}$, shows significant changes in velocity during observations (e.g. Matzeu et al. 2016, 2017). Why this would not be seen in the far more rapidly variable AGN IRAS 13224−3809 is not obvious. It is possible that there are velocity changes on timescales that we cannot resolve, so we see averaged (and therefore broadened) absorption lines, or it could instead be that there is an intrinsic difference between these outflows.

# 7 CONCLUSIONS

We fit the spectra from the 1.5 Ms *XMM-Newton* and 500 ks *NuSTAR* observing campaign on the extreme NLS1 IRAS 13224−3809 with physical broad-band models. We analyze stacked spectra, as well as flux-resolved and time-resolved spectra. Our main results are as follows:

(i) IRAS 13224−3809 is the most extremely variable AGN. The 0.3–10.0 keV band light curve shows rapid variability on timescle down to kiloseconds, and we find a peak flux 100 times the lowest level.

(ii) Two reflection components with different ionization ($\log(\xi_{1,2}/{\rm erg\,cm\,s^{-1}})$=3.13, 1.48) are required to fit the stacked spectra, as found by previous authors.

(iii) The emissivity profile of the reflected emission steepens as the powerlaw flux drops. This can be explained by a variable primary source height in the lamp post scenario.

(iv) The variable blackbody component, used to fit some of the soft excess, follows the $F \propto T^4$ relation, indicating a constant emission area in the soft band.

(v) Four blueshifted absorption lines (Ne X, S XVI, Mg XII and Si XIV) are detected in the stacked EPIC spectra. They can be fitted by two xstar absorbers, with ionization $\log(\xi'_{1,2}/{\rm erg\,cm\,s^{-1}})$=3.72,3.05 and velocity $v_{1,2} = 0.267$, 0.225 c, confirming the presence of the UFO found by Parker et al. (2017b). The inclusion of these absorption features does not have any significant impact on the relativistic blurring parameters, indicating that the measurements made using relativistic reflection are robust.

(vi) The UFO absorption lines are prominent at low flux levels (MF and LF), which may result from the increasing ionization of the gas by the increasing X-ray flux.

(vii) A high density disk model with number density $n_{\rm e} > 10^{18.7}\,{\rm cm^{-3}}$ can potentially fit the soft excess and lessen the super-solar iron abundance requirement for the reflection components.

# ACKNOWLEDGEMENTS

J.J. acknowledges support by the Cambridge Trust and the Chinese Scholarship Council Joint Scholarship Programme (201604100032). A.C.F., M.L.P. and C.P. acknowledge support by the ERC Advanced Grant 340442. B.D.M. acknowledges support by the Polish National Science Center grant Polonez 2016/21/P/ST9/04025. W.N.A. acknowledges support from the European Union Seventh Framework Program (FP7/2013–2017) under grant agreement n.312789, StrongGravity. G.M. acknowledges support from the European Union Seventh Framework Program (FP7/2007–2013) under grant agreement n.312789, StrongGravity and the Spanish grant ESP2015-65597-C4-1-R. D.J.K.B. is supported by the Science and Technology Facilities Council (STFC). J.A.G. acknowledges the support from the Alexander von Humboldt Foundation. D.R.W. is supported by NASA through Einstein Postdoctoral Fellowship grant number PF6-170160, awarded by the Chandra X-ray Center, operated by the Smithsonian Astrophysical Observatory for NASA under contract NAS8-03060. D.J.W. acknowledges support from an STFC Ernest Rutherford fellowship. Based on observations obtained with XMM-Newton, an ESA science mission with instruments and contributions directly funded by ESA

Member States and NASA. This project has made use of the Science Analysis Software (SAS), an extensive suite to process the data collected by the *XMM-Newton* observatory. This work made use of data from the *NuSTAR* mission, a project led by the California Institute of Technology, managed by the Jet Propulsion Laboratory, and funded by NASA. This research has made use of the *NuSTAR* Data Analysis Software (NuSTARDAS) jointly developed by the ASI Science Data Center and the California Institute of Technology.

This paper has been typeset from a TeX/LaTeX file prepared by the author.